# Signatures of the π-mode anomaly in (1+1) dimensional periodically-driven topological/normal insulator heterostructures

Yiming Pan[1,2], Zhaopin Chen[1,3], Bing Wang[4], Eilon Poem[1]

1. Department of Physics of Complex Systems, Weizmann Institute of Science, Rehovot 76100, ISRAEL
2. Physics Department and Solid-State Institute, Technion, Haifa 32000, ISRAEL
3. Department of Electrical Engineering Physical Electronics, Tel Aviv University, Ramat Aviv 69978, ISRAEL
4. National Laboratory of Solid State Microstructures and School of Physics, Nanjing University, Nanjing 210093, CHINA

†Correspondence and requests for materials should be addressed to Y.P. (yiming.pan@campus.technion.ac.il).

## Abstract

Akin to zero-mode anomalies, such as the chiral anomaly of edge states in quantum Hall effect, in this work, a pi-mode anomaly is proposed in a 1+1 dimensional periodically-driven topological/normal insulator (TI/NI) heterostructure. Usually, when coupling in a background gauge field, the zero modes on domain walls would provide an anomalous current term that is eventually canceled by additional boundary contributions from the topological bulk, via the Callan-Harvey mechanism. This anomaly cancellation associated with the generalization of bulk-boundary correspondence is called anomaly inflow. Through our photonic modeling and setup of the Floquet TI/NI heterostructure, for the first time, we experimentally observed the $\pi$-mode domain wall in certain driven frequencies, which is always attached to the reminiscent Floquet gauge that plays the vital role of an emergent background field. Indeed, due to the possible emergence of Floquet gauge anomaly from the driven topological bulk, the resultant $\pi$-mode anomaly can be matched on the driven interface between Floquet domains. Prospectively, we believe our prediction and observation could pave a new avenue on exploring anomalies in both periodically-driven classical and quantum systems.

Quantum anomaly is an anomalous current that violates the corresponding symmetry of classical physics at quantum level [1-3]. It offers an anomalous symmetry breaking mechanism, slightly different from spontaneous and explicit symmetry breakings [4]. Initially, the anomalous current term is discovered and discussed in the neutral pion decay process by Alder [5], Bell and Jackiw [6] in 1969, and thus, named as "Alder-Bell-Jackiw anomaly". In 1979, Fujikawa developed a heat-kernel regularization technique and sought the anomalies in its path-integral measure for gauge-invariant massless fermions [7, 8]. Later, people found a wide number of anomalies in quantum field theories. Taking Dirac fermions as an example, when interacting with a background gauge field, they give rise to chiral anomalies in even-dimensional spacetime and parity anomalies in odd-dimensional spacetime. Until the mid-1980s, we had already achieved a full understanding of anomalies by means of the Atiyah-Singer index theorem that is directly to connect it with topology [1-3], and by the Callan-Harvey mechanism that cancels the zero-mode anomalies on defects (domain walls, vortices, strings, etc.) through anomaly inflow from the extra dimension [9-12].

On the other hand, the occurrence of anomalies was further verified in the development of condensed matter physics, which is contributed to the discovery of quantum Hall effect (QHE) (1980) [10, 13-18] and sequentially that of symmetry-protected topological phases (SPT phases) [19-22]. The QHE of two-dimensional electron gas can be regarded as the material realization of parity anomaly [14, 16, 21], and Weyl semimetals proposed in TaAs family [22, 23] hold 3+1D chiral anomaly by splitting a Dirac node into two non-degenerated Weyl points in Brillouin zone that results into chiral magnetic effect [24, 25], anomalous (thermal) Hall conductivities [26-28] and negative magnetoresistance [23, 24], as exotic responses to an electromagnetic field (and temperature). Also, the time-reversal $Z_2$ SPT phases can be rephrased as a manifestation of the 't Hooft anomaly of global discrete symmetries [3, 29, 30]. Moreover, recent development about the understanding of anomaly inflow as nontrivial consequent edge states (zero modes) on the boundary or interface of SPT orders [3, 19-21], elucidates the profound connection between quantum anomalies in high-energy physics and the bulk-boundary correspondence in topological materials.

Triggered by the generalization and implementation of topological phases in periodically-driven systems [31-36], we proposed a new kind of driven-induced gauge anomaly associated with the anomalous topological phase on a Floquet domain wall, which we called "pi-mode anomaly." In this *Letter*, we found that unlike the counterpart zero-mode anomaly requiring the coupling of an external gauge field, the pi-mode anomaly arises simultaneously when the driven domain wall is constructed at appropriately intermediate frequencies of our protocols [32, 33]. In this situation, the remaining Floquet gauge associated with the micromotion in Floquet engineering is playing the background gauge field's role. Our prediction and observation unveil a new kind of Callan-Harvey mechanism that the pi-mode anomaly on domain wall could be canceled by a newly-reported gauge anomaly inflow from Floquet bulk.

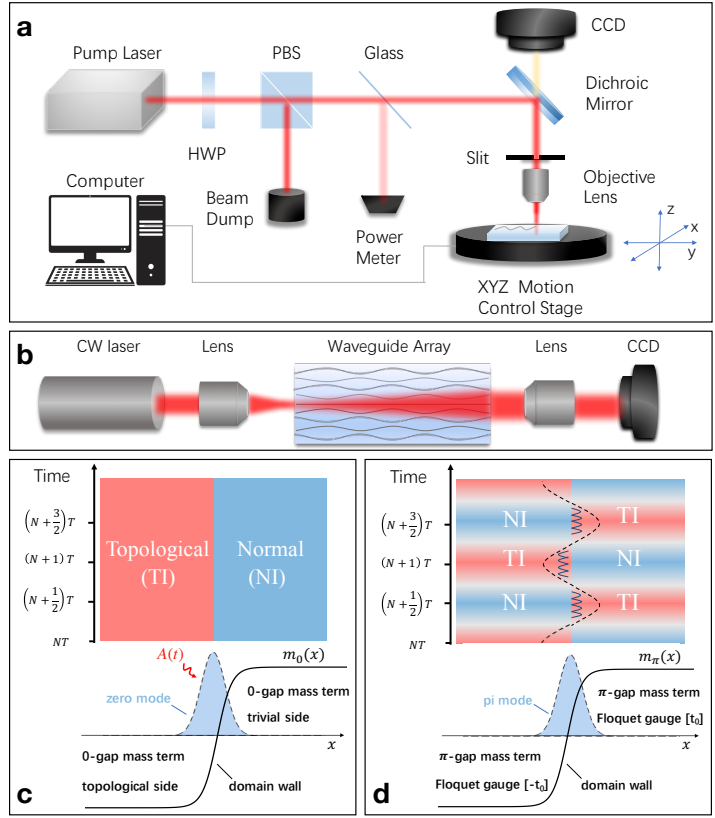

Fig. 1: The femtosecond direct-writing setup and the topological domain-wall constructions in the (1+1) dimensional periodically-driven topological insulator/normal insulator (TI/NI) heterostructures. (a) The writing system for fabricating the coupled waveguides with the controllable waveguide-curving profiles. (b) The reading system for detecting the output laser intensity distribution of the array. (c) The domain wall construction for zero modes. A 0-gap mass term along x-direction smoothly changes the sign around x = 0. (d) The anomalous domain wall construction in driven TI/NI heterostructures. The 0-mode and π-mode domain walls emerge, in response to the 0- and π-gap mass kink configurations, respectively.

*Setup and modeling* - To observe the pi-mode anomaly in a Floquet system, we construct a periodically-driven TI/NI heterostructure via the state-of-the-art femtosecond laser direct writing technique [37-41]. The direct-writing simulation setup, as shown in Fig. 1a and 1b, consists of the writing system and reading system, respectively. The details of the direct-writing setup are given in Supplementary Material (SM) file. By mimicking the light propagation along with the evanescent-coupled waveguides (the propagation direction z) with an electron wavefunction evolving in a crystal in time domain (t), we equivalently map the coupled-mode theory with the tight-binding Schrödinger equation, as a compelling methodology, to facilitate the rapid growth of topological photonics [42-44]. As a result, we can fabricate an array composed of the curved waveguides to experimentally observe the electron dynamics of Floquet TI/NI heterostructure. The driven heterostructure with the emergent pi-mode at the interface region is schematically depicted in Fig. 1b.

To characterize the existence of the pi-mode domain wall, we compare the periodically-driven TI/NI heterostructure with the conventional TI/NI heterostructure. As demonstrated in Fig. 1c, a topological insulator interfaced with a normal or regular insulator, forming a heterojunction independent with time, is well understood via bulk-boundary correspondence and domain walls (also known as kink, soliton) [10, 12]. By contrast, we construct a periodically-driven TI/NI heterostructure (Fig. 1d). The driven heterostructure is composed of two distinct topological phases, but these two phases are periodically interchanged their identities in time domain. Analytically, to describe the driven TI/NI heterostructure, we assume that the left-handed side of the heterostructure system is defined by

$$H^{(L)}(t) = \begin{cases} H_{TI}, & t \in \left(-\frac{T}{4}, \frac{T}{4}\right), \\ H_{NI}, & t \in \left(\frac{T}{4}, \frac{3T}{4}\right). \end{cases} \quad (1)$$

and the right-handed side

$$H^{(R)}(t) = H^{(L)}(t + T/2), \quad (2)$$

in which the topological invariants (e.g., Chern number, Zak phase, and Pfaffian) for the two instantaneous bulk Hamiltonians ($H_{TI}, H_{NI}$) are different,

$$\nu(H_{TI}) \neq \nu(H_{NI}). \quad (3)$$

Atiyah-Singer theory tells us that the index of the Dirac operator on domain wall equals to the difference between the topological numbers of two sides [2, 19], and for the static TI/NI heterostructure system (Fig. 1c) the index equals to the number of zero modes appearing at the interface. Considering the bulk Hamiltonian (1) are time-periodic, we have to extend the definitions of topological invariants into periodically-driven systems [31].

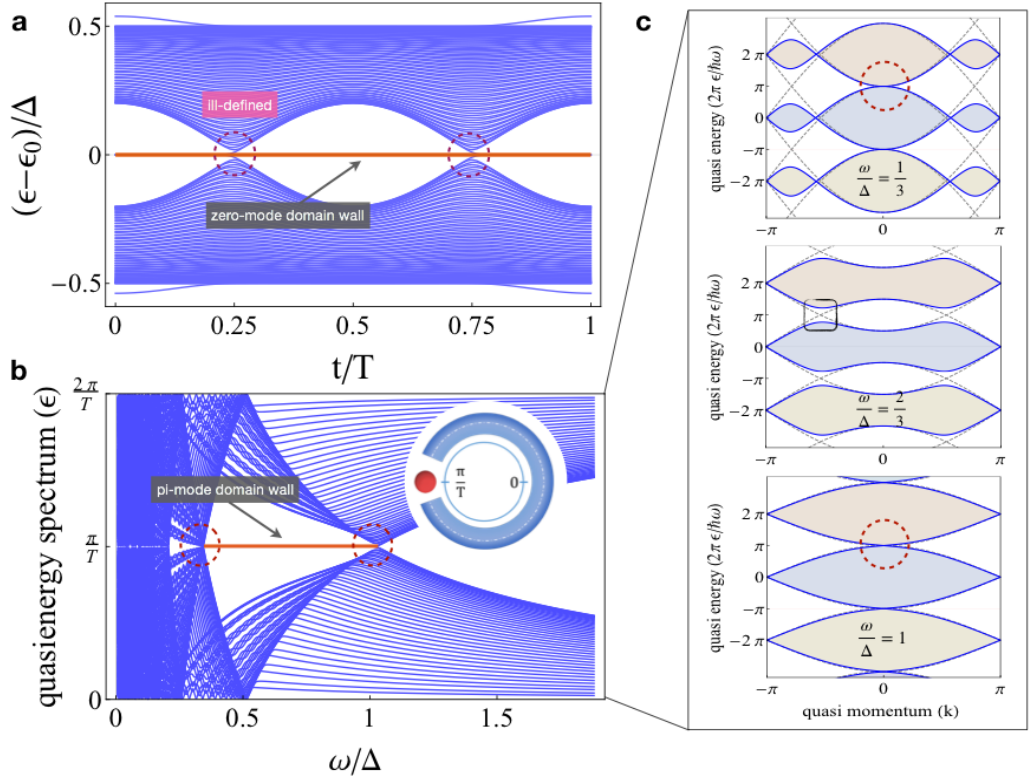

Fig. 2: The energy spectrum in the adiabatic limit ($\omega \to 0$) and the quasienergy spectrum from Floquet-Bloch theorem. (a) The invariant of the instantaneous 0-mode domain wall is ill-defined at the critical transition point when the normal phase switches to the topological phase periodically, in which the 0-gap has to be closed. (b) The emergence of the pi-mode domain wall in the quasi-energy spectrum. (c) The "open-close-open" mechanism of topological $\pi$ gap in quasi-energy-momentum space. At the driven frequencies $\omega = 1/3, 1$, the $\pi$-gap is closed nontrivially. The relevant parameters in the unit of bandwidth $\Delta$ are $\kappa_0 = 0.25, \delta\kappa_1 = 0.2, N = 19$.

*The π-mode domain wall in driven SSH setup* - Let us take a specific example, the one-dimensional driven Su-Schrieffer-Heeger (SSH) model [45, 46]. We can define the bulk

Hamiltonian and the domain wall structure via the dimerized coupling profiles, which is given by

$$H^{(L,R)}(t) = \sum_{i=1}^{N-1} \left[ \kappa_0 + (-1)^i \delta\kappa_1 \cos(\omega t + \theta^{(L,R)}) \right] c_i^\dagger c_{i+1} + h.c., \tag{4}$$

where $c_i^\dagger$ and $c_i$ are the creation and annihilation operators at the site i (or the i$^{th}$ waveguide), N is the total number of lattice sites. The second off-diagonal term in Eq. (1) represents the nearest-neighbor hopping, in which $\kappa_0$ is the constant coupling strength and $\delta\kappa_1$ is the amplitude of the periodically dimerized staggered coupling strength with $\omega = 2\pi/T$ being the driven frequency and θ the initial phase of the drive.

To explicitly explore the gap opening mechanism of the isolated modes, let us represent our driven SSH Hamiltonian on the Bloch basis. Considering the periodic boundary condition, due to the translation symmetry, we can transform the equation (4) into the momentum space with the corresponding Bloch representation as [36, 48] $H^{(L,R)}(k,t) = \left(\kappa_0 - \delta\kappa_1 \cos(\omega t + \theta^{(L,R)})\right) + \left(\kappa_0 + \delta\kappa_1 \cos(\omega t + \theta^{(L,R)})\right) \cos(k))\sigma_x + \left(\kappa_0 + \delta\kappa_1 \cos(\omega t + \theta^{(L,R)})\right) \sin(k) \sigma_y$, where $\sigma_x, \sigma_y$ are the Pauli matrices on the basis of sublattices A and B, and k is the momentum index with the lattice constant a = 1 for the dimerized supercell.

Two reasons of our choice are worth mentioning here. First, the driven SSH model is simple to realize by employing the laser direct-writing technique. We can fabricate the curved waveguide array to observe the anomalous behaviors in Floquet systems. Second, the SSH model is one of the stereotypical one-dimensional topological insulators that equivalently connects with Kitaev's toy model for a spinless p-wave superconductor and the Ising model in a transverse field for spin chains [47]. Accordingly, the resultant pi-mode anomaly in our experiment can be easily implemented in generic one-dimensional symmetry-protected Floquet topological phases.

In the context of Floquet engineering [32, 33], the choice of the initial phase $\theta$ is associated with the initial time $t_0$, which is termed as Floquet gauge. For our concern, we choose a particular gauge $\theta^{(L)} = 0$ for H$^{(L)}$ and $\theta^{(R)} = \pi$ for H$^{(R)}$, respectively, in order to satisfy the kink setup condition (2). In general, the choice of Floquet gauge, namely, the choice of $\theta$, or $t_0$, affects the form of the Floquet Hamiltonian describing the stroboscopic dynamics. The stroboscopic Floquet Hamiltonian is given by

$$H_F[t_0] \equiv \frac{i}{T} \ln\left(\widehat{T} e^{-i \int_{t_0}^{t_0+T} H(t')dt'}\right), \quad (5)$$

where the periodicity of the driven bulk Hamiltonian is denoted as $H(t) = H(t + T)$, and as expected, the Floquet gauge is periodic and continuous, $H_F[t_0] = H_F[t_0 + T]$. Thus, we find that the corresponding Floquet Hamiltonian of both sides of the bulk system have the relation, $H_F^{(L)}[t_0] = H_F^{(R)}[t_0 + T/2]$, or equivalently $H_F^{(L)}[\theta] = H_F^{(R)}[\theta + \pi]$.

Before we resolve the Floquet system with the quasienergy spectrum, let us first take into account the two limits of Floquet analysis: the adiabatic limit ($T \to \infty$) and the high-frequency limit ($T \to 0$). First, in the low-frequency regime, we can directly calculate the eigenvalue spectrum of the instantaneous Hamiltonian $H(t) = H^{(L)}(t) + H^{(R)}(t)$ as a function of time t in a driven period T, as demonstrated in Fig. 2a. We observe that an instantaneous zero-energy mode exists in a fully driven cycle, implying the presence of zero-mode domain wall at the interface between the TI/NI junction. However, there is a problem that the 0-gap closes at two critical times $t = T/4$ and $3T/4$. It leads to the critical phase transition when the driven staggered coupling disappears instantly, for the reason that the gap has to be closed when a normal insulating phase continuously becomes a nontrivial insulating phase with a distinct topological number. Consequently, the instantaneous Hamiltonian is forced to be metallic at the critical point (see the massless Dirac equation in Fig. S3f of the SM file). The 0-gap invariant is ill-defined in the driven TI/NI heterostructure. Thus, the instantaneous zero-mode domain wall loses its topological protection and quickly being scattered into the bulk. For the driven TI/NI heterostructure setup, the underlying principle of bulk-boundary correspondence is violated in the adiabatic limit.

Second, in the high-frequency limit, we can achieve the effective high-frequency-approximated Hamiltonian [33], which is obtained as

$$H_{\text{eff}}^{(L)} = \lim_{T \to 0}\left(\frac{1}{T}\int_0^T H^{(L)}(t)dt\right) = \lim_{T \to 0}\left(\frac{1}{T}\int_0^T H^{(R)}(t)dt\right) = H_{\text{eff}}^{(R)}, \quad (6)$$

Thus, the interface of the driven TI/NI heterostructure disappears in the high-frequency approximation, and equivalently there is no domain wall. Indeed, the explicit Magnus expansion is given by $H_{eff}^{(L,R)} = H_0 + O(\frac{1}{\omega^n})$, and in the limit ω → ∞, it enables us to ignore the high-order contributions [32]. Besides, the zero-order term $H_0 = \sum_{i=1}^{N-1} \kappa_0 c_i^\dagger c_{i+1} + h.c.$ is the same as the metallic Hamiltonian (4) without dimerization, in accordance with the massless Dirac equation in continuum limit (Fig. S3c and S3f).

Finally, since both the adiabatic and high-frequency limits of the driven heterostructure are trivial, we, therefore, explore the intermediate resonant frequency regime, i.e., ω~Δ, with Δ=||$H_0$||=4$\kappa_0$ being the static bandwidth. Fig. 2b demonstrates the quasienergy spectrum by directly solving the eigenvalue problem from the Floquet Hamiltonian $H_F[t_0]$ (5) (the calculation can be seen in the SM file). From the Floquet quasienergy spectrum, we observe that the pi-mode domain wall survives at a certain frequency range $\frac{\omega}{\Delta} \in \left(\frac{1}{3}, 1\right)$. It means that instead of opening the 0-gap, the π-gap is nontrivially opened, allowing the occurrence of the pi-mode domain wall in our Floquet setup.

Three spectral features are worth mentioning here. (i) From the quasienergy spectrum, we can easily check that the adiabatic and high-frequency regions are trivial, consistent with our previous analysis. (ii) The critical frequencies 1/3 and 1 can be explained by the level crossing and avoiding between the artificial photon bands (Floquet replica) in momentum space [36], as shown in Fig. 2c. The π-gap is closed at $\frac{\omega}{\Delta} = \frac{1}{3}$ due to the touching between the Floquet replica $(n + 1)$ and $(n - 2)$ and meanwhile closed at $\frac{\omega}{\Delta} = 1$ due to the touching between replica $n$ and $(n + 1)$. Whereas in the range $\frac{1}{3} < \frac{\omega}{\Delta} < 1$, only the level crossing between the neighboring replica and the scattering is allowed because the non-zero contribution ($H^{(\pm 1)} \neq 0$). The periodic staggered coupling strength gives rise to the level avoiding at the quasienergy π/T, indicating that it opens the π-gap, rather than the 0-gap. (iii) The π-gap invariant $\nu_\pi$ is explicitly given in the literature [34, 35]. We calculate the nontrivial gap invariant separately for the left-handed and right-handed Hamiltonians, and they are equal, $\nu_\pi(H^{(L)}) = \nu_\pi(H^{(R)}) = \frac{i}{2\pi} \int_{BZ} Tr((V_\pi^+)^{-1} \partial_k V_\pi^+)$, where $V_\pi^+$ corresponds to the periodized evoluation operator (see the calcualtion in the SM file) [34]. The naïve implementation of Atiyah-Singer index theorem implies no domain wall fermions due to $|\nu_\pi(H^{(L)}) - \nu_\pi(H^{(R)})| = 0$. Paradoxically, we indeed achieve the pi-mode domain wall in the specific driven frequencies (Fig. 2b).

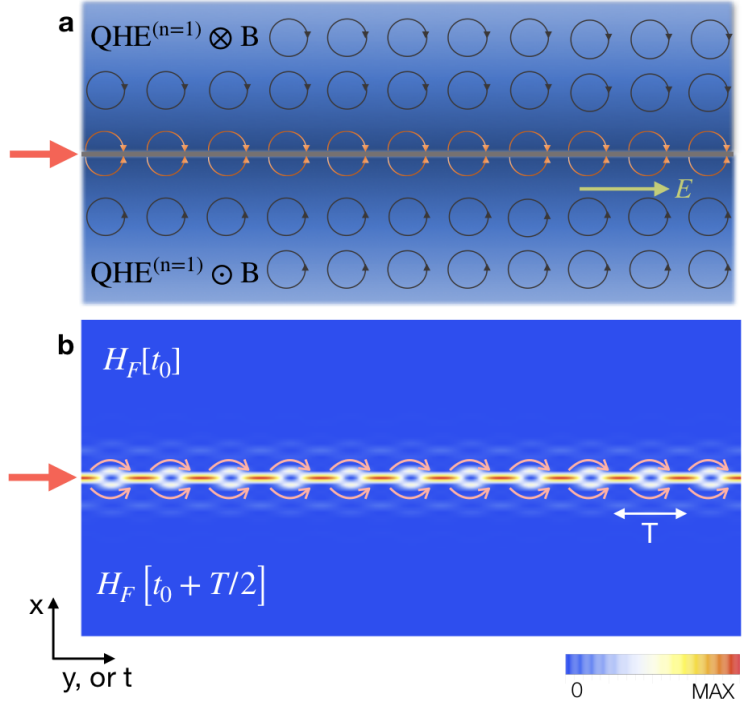

Fig. 3: Comparison between zero-mode anomaly and pi-mode anomaly. (a) The gauge anomaly of chiral edge states on the domain-wall defect of the QHE heterostructures with the filled Landau level $n = 1$ and the opposite magnetic fields exerted on both sides. (b) pi-mode anomaly on a driven domain wall in the time-periodic TI/NI heterostructures, as constructed in Fig. 1d. Notice that the Floquet Hamiltonian is gauge-dependent $H_F[t_0] = H_F[t_0 + T]$. The relevant numerical parameters in the unit of bandwidth $\Delta$ are $\kappa_0 = 0.25, \delta\kappa_1 = 0.2, N = 19, \omega/\Delta = 0.75$.

*Floquet gauge anomaly* – To remedy the contradiction, we should take a close look at the associated Floquet gauge $t_0$. Even though the gap invariants of both lopsided bulk phases of the TI/NI heterostructure are the same, the pi-gaps are intimately tied with different Floquet gauge choices due to the relation (2). The pi-gap can be denoted as the mass term for the massive Dirac equation: $m_\pi^{(R)}(t_0) = m_\pi^{(L)}(t_0 + T/2)$ and by considering the linearity with the staggered coupling strength in (4), we can find that $m_\pi^{(L)}(t_0 + T/2) = -m_\pi^{(L)}(t_0)$ at the case $\theta = \pi$. Thus, the two bulk Floquet Hamiltonians on the left-handed and right-handed sides have a pi-gap mass term with opposite sign inherited from the Floquet gauges. Consequently, the pi-mode domain wall should emerge as a result of the genuine non-zero index $\left|v_\pi(H^{(L)}) - (-v_\pi(H^{(R)}))\right| = 2$ [51].

Moreover, we realize that the dynamics of the pi-mode domain wall can be assumed as the intrinsic pi-mode anomalous term on the interface that is compensated by the Floquet gauge anomaly from the Floquet bulk [30]. As demonstrated in Fig. 3, we compare a prototypical zero-mode anomaly in a quantum hall heterostructure with the pi-mode anomaly in our driven TI/NI heterostructure. A 2D free electron gas (x-y plane) is driven by an external magnetic field B in its perpendicular direction (z) to form a quantum Hall effect (QHE). This QHE ground state can be described by the Chern-Simons theories [13,19-21]. We construct a QHE heterostructure by switching the magnetic field directions in the upper half-plane and lower half-plane, as shown in Fig. 3a. By adding the electric field E in the y-direction, we observe two currents flow: the first is the anomaly inflow from the QHE bulk toward the interface of the heterostructure, which is the anomalous boundary term from the Chern-Simons and known as the transverse Hall current [10, 13]; the second is that the resulting chiral edge states on the domain wall when interacting with the parallel electric field ($E_y$), gives rise to the gauge anomaly along the y-direction. The anomalous boundary term from the Chern-Simons bulk, also known as parity anomaly [16], can exactly match the zero-mode anomaly on the interface, yielding the preservation of gauge symmetry. This anomaly cancellation is called as the Callan-Harvey anomaly inflow mechanism [10, 12].

To evaluate the pi-mode anomaly, we should decompose the dynamical evolution in a Floquet system into two components [32, 33, 42]: the stroboscopic evolution that leads to the anomalous Floquet topological phases (T); the micromotion that plays the role of an effective or "inertial" background gauge field ($t_0$). The decomposition of the generic time evolution operator is given by

$$U(t, t_0) = U(t, t_0 + nT)[U(t_0 + T, t_0)]^n = V(t, t_0) e^{-iH_F[t_0](t-t_0)}, \qquad (7)$$

where the periodized evolution operator is defined as $V(t, t_0) \equiv U(t, t_0) e^{iH_F[t_0](t-t_0)}$ that contains the short-time scale information and topological invariants (namely, the micromotion), while the Floquet Hamiltonian $H_F[t_0]$ (5) contains the long-timescale information and non-equilibrium steady states (namely, the stroboscopic dynamics). Now we can define a Floquet-gauge-independent Floquet Hamiltonian $H_F$ through the periodized evolution operator [32],

$$H_F[t_0] = V^{-1}(t,t_0)(H(t) - i\partial t)V(t,t_0) = V^{-1}(t,t_0)H_F V(t,t_0), \tag{8}$$

where we find that the effect gauge-independent Hamiltonian $H_F = H(t) - i\partial t$ is exactly the differential operator for the original time-periodic Schrödinger equation ($i\partial t \psi(t) = H(t)\psi(t)$). In light of the relationship between the equivalent Floquet Hamiltonian (8), we can denote the gauge-independent $H_F$ as "the differential form", and the gauge-dependent $H_F[t_0]$ as "the integral form" [32, 33]. More importantly, we can define a Floquet-gauge-associated kick operator $K(t,t_0) \equiv \frac{i}{T} \ln V(t,t_0)$. It connects to the micromotion dynamics within a Floquet cycle for the specified initial value of $t_0$, acting as the background gauge field to the stroboscopic dynamics of the Floquet Hamiltonian. To further assess the presence of Floquet gauge in π-gap mass term, we reduce the Floquet Hamiltonian into two relevant Floquet replica between n=1 and n=0 with the $k \cdot p$ approximation (see the middle subfigure in Fig. 2c)

$$H_{FD}^{(\pi)}[t_0] = \frac{\omega}{2}I_2 \pm \sqrt{1 - \left(\frac{\omega}{\Delta}\right)^2} \begin{pmatrix} \kappa_0 p & ie^{i\omega t_0}\delta\kappa_1 \\ -ie^{-i\omega t_0}\delta\kappa_1 & -\kappa_0 p \end{pmatrix}, \tag{9}$$

where $p = k \pm \arccos(\omega/\Delta)$ and $\Delta = 4\kappa_0$. The local Floquet gauge for left- and right-handed Floquet bulk can be rephrased as $\theta^{(L)} = \omega t_0 \ (x < 0), \theta^{(R)} = \omega\left(t_0 + \frac{T}{2}\right)(x > 0)$. The corresponding gauge-independent Floquet-Dirac Hamiltonian is given by $H_{FD}^{(\pi)} = V(t_0)H_{FD}^{(\pi)}[t_0]V^{-1}(t_0) = \frac{\omega}{2}I_2 \pm \sqrt{1 - \left(\frac{\omega}{\Delta}\right)^2}(\kappa_0 p \sigma_z - \delta\kappa_1 \sigma_y)$ with a unitary transformation $V(t_0) = e^{-i\omega t_0 \sigma_z/2}$. Therefore, we can define a Floquet gauge parity ($Z_2$) that relates to the micromotion operator $V(t_0) = V(t_0 + T), V(t_0 + T/2) = -V(t_0)$. The Floquet-Dirac mass term $m_\pi = \sqrt{1 - \left(\frac{\omega}{\Delta}\right)^2} \delta\kappa_1 \sigma_y$ is gauge-independent. Employing the above micromotion operator to mass term separately, we can obtain

$$\begin{aligned} m_\pi^{(L)} &= V^{-1}(t_0)m_\pi V(t_0) = \sqrt{1 - \left(\frac{\omega}{\Delta}\right)^2} \delta\kappa_1 e^{-i\omega t_0 \sigma_z} \sigma_y, \\ m_\pi^{(R)} &= V^{-1}(t_0 + T/2)m_\pi V(t_0 + T/2) = -\sqrt{1 - \left(\frac{\omega}{\Delta}\right)^2} \delta\kappa_1 e^{-i\omega t_0 \sigma_z} \sigma_y. \end{aligned} \tag{10}$$

Thus, it indicates that the left-handed and the right-handed mass terms are parity-odd, i.e., $m_\pi^{(R)} = -m_\pi^{(L)}$, convincing a 't Hooft-like anomaly on the driven wall associated with a $Z_2$ symmetry [3, 29, 30, 49]. To demonstrate the anomaly inflow associated with the spatially-twisted Floquet gauge (Fig. 3b), alternatively, we apply the Glodstone-Wilczek approach [10,11] to calculate the current of Floquet-Dirac Hamiltonian (9) induced by the twisted gauge configuration $t_0(x)$, which is obtained as

$$\langle j^\mu \rangle = \frac{1}{2\pi} \epsilon^{\mu\nu} \partial_\nu \big(\omega t_0(x)\big), \tag{11}$$

where in 1+1 spacetime $\mu, \nu = 0,1$, $\epsilon^{\mu\nu}$ is an anti-symmetry tensor with $\epsilon^{01} = 1$. The total quantum number is then given by

$$Q_\pi = \int j^0 dx = \frac{\omega}{2\pi} \int \partial_x t_0(x) dx = \frac{\omega}{2\pi} t_0(x)\big|_{x=-\infty}^{x=+\infty} = \frac{\omega}{2\pi}\left(t_0 + \frac{T}{2} - t_0\right) = \frac{1}{2}. \tag{12}$$

The fractional charge $Q_\pi = 1/2$ is exactly the soliton fermion number of the static Jackiw-Rebbi model [11, 17, 45, 50, 51]. As demonstrated in Fig. 3b, we can observe the stroboscopic eigenstate dynamics of pi-mode in time domain mimicking as a current term (Eq. 11) flowing inward the interface. This is the expected "Floquet gauge anomaly", inspired by the Callan-Harvey anomaly inflow mechanism [11], which explicitly matches the pi-mode anomaly on the driven wall. In addition, we notice that the driven fractional charge can be arbitrary such as $Q_\pi = 1/N$ if the right-handed Floquet gauge is relatively shifted to $t_0 + \frac{T}{N}$.

Comparing the zero-mode anomaly in the 2D QHE heterostructure (Fig. 3a) and the pi-mode anomaly in the 1+1D driven TI/NI heterostructure (Fig. 3b), we conclude three aspects to address the similarities, differences and relations of the two anomalies. The first is the dimensionality of the two anomalies. The zero-model anomaly is 1+1D chiral anomaly [9, 10], in which the anomalous current flows along the spatial dimension (y), and the temporal dimension is a dummy variable. In contrast, the dimension of the pi-mode anomaly is 0+1D where the temporal dynamics in time domain (t) is relevant since the spatial dimension is zero. Nevertheless, our driven TI/NI heterostructure is easily implemented in higher-dimensional Floquet systems and at the moment the pi-mode anomaly behaves as the anomalous photoinduced current at the interface [51]. The second is that the zero-mode anomaly is induced by adding an external gauge field to the zero modes and an additional spatially-modulated mass term is needed to support domain-wall fermions [10], whereas the pi-mode anomaly is intrinsically achieved because the protocols of Floquet engineering give arise to both the pi-gap mass term in stroboscopic evolution and the emergent Floquet gauge in micromotion simultaneously. The last is that the anomaly cancellation mechanism for the zero-mode domain-wall fermions holding the global anomalies or gauge anomalies has been well investigated in quantum path-integral measure [1, 10, 49]. However, the quantum-field description for pi-mode domain walls and pi-mode anomalies in periodically-driven anomalous topological phases is still unknown and challenging.

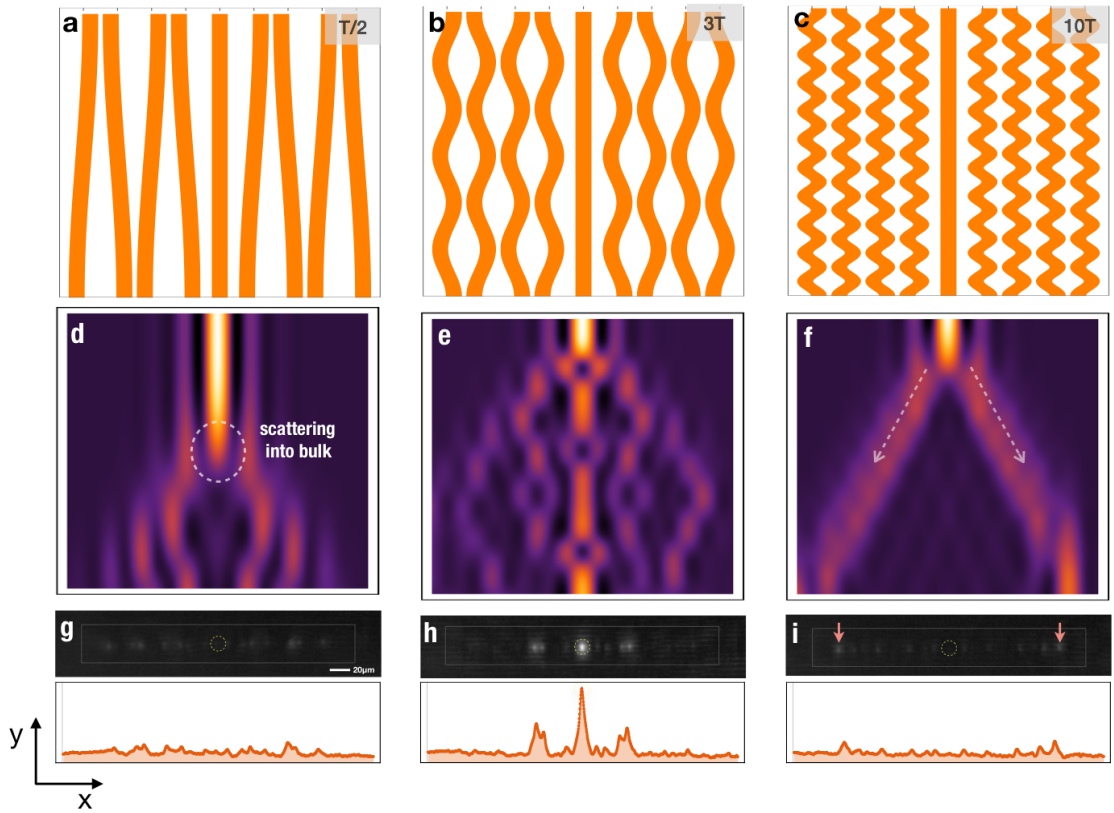

Fig. 4: Experimental observations of driven domain wall dynamics ranging from adiabatic limit (a) to high-frequency limit (c). The pi-mode anomaly on the domain wall appears at a specific driven frequency range (b, e, h), with the curving period T = 8.3 mm, where the π-gap invariant is nontrivial in quasienergy spectrum (see Fig. 2b). The fabrication (a,b,c), simulation (d,e,f), experiment (g,h,i) parameters are presented in the main text.

*Experimental observation* – To experimentally verify the pi-mode anomaly in our designed driven TI/NI heterostructure, we fabricated the corresponding coupled curved waveguides through the femtosecond direct-writing method and measured its output intensity distribution (Fig. 1b). Fig. 4 demonstrates the array fabrication, theoretical prediction, and the output measurement in terms of the different Floquet periods (T), such that the adiabatic regime, the intermediate driven regime, and the high-frequency regime. The waveguide number of the array is N=19, the propagation length is L=25 mm, and the extremal staggered waveguide spacings are 10 μm and 20 μm. Notice that the center waveguide (the $10^{th}$) for the light input is straight. The periodicity of driven forces is imprinted on the curving profiles of waveguides, leading to the time-periodic spacing between two given waveguides. As we experimentally extracted (see Fig. S2 in the SM file), these two staggered spacings correspond to the instantanous minimal and maximal coupling strengths $0.544\ mm^{-1}$ and $0.071\ mm^{-1}$, which leads to the linearly-averaged coupling strength $\kappa_0 = 0.3\ mm^{-1}$. Taking into account the driven condition $1/3 < \omega/4\kappa_0 < 1$, for our setup we can estimate the pertinent curving period $5.2\ mm < T < 15.7\ mm$.

As we compared with the three driven regimes, only the output intensity in the intermediate regime manifests the localization property of the output intensty distribution in the center waveguide of the array, as shown in Fig. 4h, with the curving period T=8.3 mm in the driven condition. The output intensities of arrays show diffusive characterization in both the adiabatic (T=50 mm) and high-frequency regimes (T=2.5 mm), in consistent with the expectation from quasienergy spectrum (Fig. 2b). The dimensionless numerical parameters are $\kappa_0 = 0.68, \delta\kappa_1 = 0.25$, N=19 and the periods are $T = 30, 8, 1.5$, respectively. As a calibration, we also fabricated the straight waveguide array based on the kink structures of the static SSH model and observe the zero-mode domain walls experimentally, see the comparison bewteen pi modes and zero modes domain wall in the SM file.

In addition, two aspects of our observation are addressed here. Firstly, our experiment is based on photonic simulation by means of mapping the genuine time (t) into the propagation direction (y) along with the waveguide array. Within this in mind, our optical measurement is an anology of the dynamics of the pi-mode domain wall in periodically-driven systems [49]. Secondly, to further detect the instantaneous micromotion dynamics inside the array heterostructure (see Fig. 3b and 4e), we demand the optical near-field measurement [36] to record the near-field intensity distribution in full propagation regime, while our reading system cannot detect the near-field evolutions. There are several related experimental methods and platforms that are able to explore the dynamics of the pi-mode anomaly [42-44]. Also, the driven TI/NI heterostructure can be implemented in other quantum simulation platforms, such as ultracold atoms, and trapped ions.

*Conclusion* – In short, as compared with zero-mode domain wall and zero-mode anomaly from quantum Hall effect, we firstly proposed a periodically-driven topological/normal insulator heterostructure to construct the pi-mode domain wall and pi-mode anomaly in Floquet systems. The pi-mode anomaly is canceled by the Floquet gauge anomaly and therefore reveals a Floquet-engineered version of the Callan-Harvey anomaly inflow mechanism. Via the femetosecond laser direct-writing technique, we experimentally observed the pi-mode anomaly and the driven domain wall in our designed curved waveguide arrays. Still, many pertinent issues have not been explored yet, such as the coexistence of 0 and pi-mode anomalies, and the anomalous bulk-boundary correspondence in generic quantum and classical Floquet systems.

**Acknowledgments**

We acknowledge Yehonatan Gilead, Uri Levy, Xinbin Song, Nir Davidson, Chenjie Wang for useful discussions and comments. In particular, we would like to thanks the deceased professor Yaron Silberberg for his inspiration and early contribution of this work. The work was supported in parts by DIP (German-Israeli Project Cooperation) No. 04340302000, ISF (Israel Science Foundation) No. 00010001000, and by ICORE— Israel Center of Research Excellence program of the ISF, and by the Crown Photonics Center.

Y. P. and Z. C. contribute equally to this work.

# Supplementary Materials:

# Signatures of the π-mode anomaly in (1+1) dimensional periodically-driven topological/normal insulator heterostructures

Yiming Pan[1,2], Zhaopin Chen[1,3], Bing Wang[4], Eilon Poem[1]

5. Department of Physics of Complex Systems, Weizmann Institute of Science, Rehovot 76100, ISRAEL
6. Physics Department and Solid State Institute, Technion, Haifa 32000, ISRAEL
7. Department of Electrical Engineering Physical Electronics, Tel Aviv University, Ramat Aviv 69978, ISRAEL
8. National Laboratory of Solid State Microstructures and School of Physics, Nanjing University, Nanjing 210093, CHINA

## 1. Waveguide fabrication and measurement

The experimental setup, including the fabrication part and the measurement part, is presented in Fig. 1a and 1b, respectively. The evanescently-coupled waveguides are fabricated in bulk glass using the femtosecond-laser direct-write technique. The laser pulse is of 500 fs-long, 12 nJ at 500 kHz repetition (with a pulse divider of 2), with a central wavelength of 1041 nm. In order to suppress the polarization dependence of the coupling, the pulse is shaped by a slit. Then it is focused on a standard 25x, NA=0.5 microscope objective into a Corning 0215 soda-lime glass with a size $25\ mm \times 25\ mm$, to produce a change of the refractive index along a designed line, a waveguide. We place the focus approximately 117 μm below the glass surface, leading to waveguide with a mode of approximately three microns FWHM. The glass slide is fixed on a computer-controlled stage, which moved at a speed of 0.2 mm per second along the y-direction. In the sample fabrication setup in Fig. 1a, HWP is the half-wave plate, and PBS is the polarizing beam splitter.

The typical example of the waveguide array is shown in Fig. 1b and Fig.4 in the main text. As a sample can be seen in Fig. S1, the fabricated waveguides start as a symmetric distribution of the Su-Schrieffer-Heeger (SSH) model with a domain wall defect, a straight waveguide, in the middle. The minimum and maximum distance of the neighbored waveguides are $10\ \mu m$ and $20\ \mu m$, respectively. The distance between the middle waveguide and its neighbored ones is also $20\ \mu m$ at the beginning. In the z-axis direction, the waveguides are designed as a periodic

Floquet structure. The neighbored waveguides, except the middle one, oscillate periodically with an amplitude of 2.5 $\mu m$, but in the opposite direction. In the fabricated samples (see Fig.4), we design three structures with 0.5, 3, and 10 Floquet cycles in the total propagation length of 25 mm. In the experimental samples, the coupled waveguide number for each waveguide array is 19. The curved waveguides are symmetrically distributed at two sides of the straight waveguide. The relation between the coupling distance and the coupling amplitude is shown in Fig. S2, the data of which is extracted from Ref. [39], as they are fabricated in the same writing system.

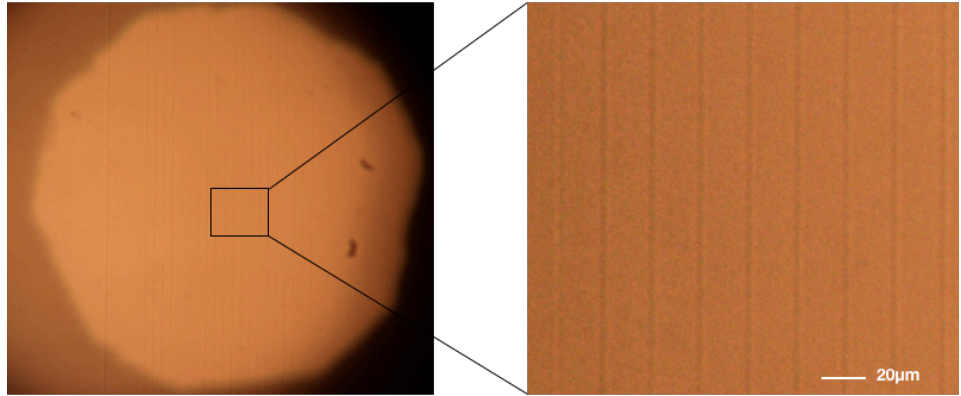

Fig. S1: the typical structures of fabricated waveguides arrays of the soda-lime glass sample in optical microscopy.

The fitting curve shows that the neighboring distance of 10 $\mu m$ and 20 $\mu m$ correspond to 0.544 $mm^{-1}$ and 0.071 $mm^{-1}$, respectively. Since the average distance between the neighboring waveguides in our samples is 15 $\mu m$, the corresponding coupling amplitude is linearly-approximated as $\kappa_0 = (0.544\ mm^{-1} + 0.071\ mm^{-1})/2 \approx 0.31\ mm^{-1}$ and the static bandwidth is given by $\Delta = 4\kappa_0 = 1.24\ mm^{-1}$. Then the driven frequency over the static bandwidth for the three experimental samples, the curving period T=50, 8.3 and 10 $mm$, are $\omega/\Delta$ =0.10, 0.61 and 2.03, respectively. Obviously, the normalized driven frequency for the middle case is located at the above-mentioned pi-mode region $1/3 < \omega/\Delta < 1$ in Fig. 2b, showing that our experimental results well match the theoretical analysis.

The reading part of the experimental setup is shown in Fig. 1b of the main text. A continuous-wave NIR laser diode is used to focus light into the waveguide. The laser light is initially coupled into the middle straight waveguide by a focus lens. Then, the output signal of intensity distribution is read out by a CCD camera.

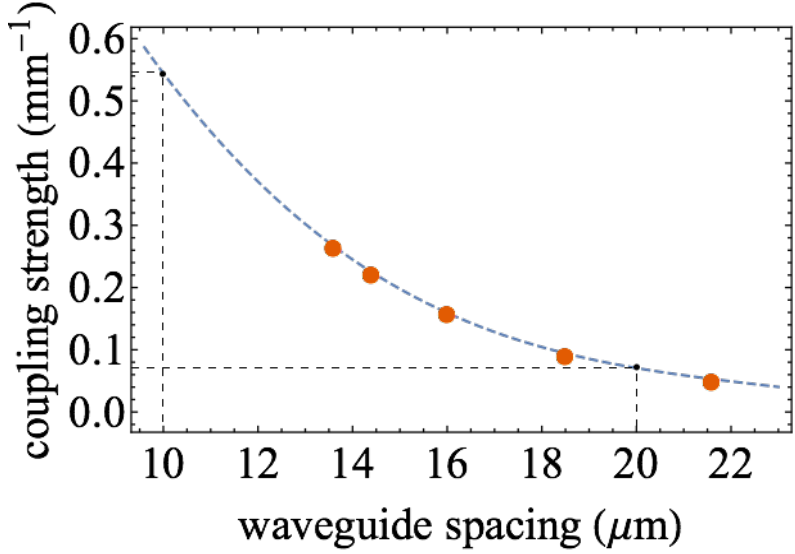

Fig. S2: the coupling configuration as a function of the distance between neighboring waveguides. Here, the fitting curve shows that the distances 10 $\mu m$ and 20 $\mu m$ correspond to 0.544 $mm^{-1}$ and 0.071 $mm^{-1}$, respectively. The data in the plot is cited from ref. [39].

**2. Domain walls in the static massive Dirac equations**

As a trial, we start with a massive Dirac Hamiltonian

$$H(x) = -i\hbar\, \partial_x \sigma_1 + m(x)\sigma_3, \qquad (S1)$$

with the kink-loaded mass term

$$m(x) = \begin{cases} -m, & x \to -\infty, \\ +m, & x \to +\infty. \end{cases} \qquad (S2)$$

Considering the zero-energy solution of the time-independent Schrodinger equation at $E = 0$, the eigenvalue equation has the form

$$[-i\hbar\, \partial_x \sigma_1 + m(x)\sigma_3]\psi^{(0)}(x) = 0,$$

multiplying $\sigma_1$ from the left-handed side, we have

$$\frac{\partial \psi^{(0)}(x)}{\partial x} = -\frac{m(x)}{\hbar}\sigma_2 \psi^{(0)}(x).$$

Thus, the corresponding wavefunction should consist of the spinor component that is the eigenstate of $\sigma_2$

$$\sigma_2 \eta_\pm = \pm \eta_\pm, \eta_\pm = \frac{1}{\sqrt{2}} \begin{pmatrix} 1 \\ \pm i \end{pmatrix}. \tag{S3}$$

So, the generic wavefunction is obtained as

$$\psi^{(0)}(x) = \frac{1}{\sqrt{2}} \begin{pmatrix} 1 \\ \pm i \end{pmatrix} \exp\left(\mp \int_0^x \frac{m(x')}{\hbar} dx'\right). \tag{S4}$$

Depending on the mass term configuration $m(\pm\infty)$, we can find that only one of zero modes with a definite chirality ($\eta_+$) is normalizable, the rest one diverges exponentially in both directions. Therefore, we obtain a chiral zero-mode that lives on the domain wall, as shown in Fig. 1c of the main text.

To stay in a 1+1D setup of the domain wall, now we will study a 2+1D heterostructure bulk system. Along the spatial dimension denoted by x, a mass defect is introduced as same as Eq. (S2), of which the explicit form of the mass distribution is not relevant. In this situation, there should provide solutions $\psi_\pm$ which are zero energy eigenstates of the Dirac Hamiltonian

$$H(x) = -\sigma_1 [\sigma_2 \partial_y + \sigma_3 \partial_x - m(x)] \tag{S5}$$

where the generic solutions are given by

$$\psi_\pm = \eta_\pm \psi_\pm^{(0)}(x) e^{ip_y y}, \tag{S6}$$

with

$$\sigma_3 \eta_\pm = \pm \eta_\pm,$$
$$\psi_\pm^{(0)}(x) = \exp\left(\mp \int_0^x \frac{m(x')}{\hbar} dx'\right)$$

Only the eigenfunction $\psi_+^{(0)}(x)$ corresponds to a normalizable solution, describing a chiral zero mode ($E = p_y$) traveling along with the interface (along the y-direction).

## 3. Domain walls in Floquet-Dirac equations with a driven-induced mass term.

To combine with the Floquet engineering, we should reassess the roles of both the gauge field and Dirac mass term playing in one-dimensional periodically-driven quantum systems. We can define a Floquet Dirac mass that is time-periodic

$$m_{FD}(x,t) = m_{FD}(x, t+T) = e\gamma_\mu A^\mu - m(x). \quad (S7)$$

Thus, we obtain a Floquet-Dirac Lagrangian

$$L = \bar{\psi}\left(i\gamma_\mu \partial^\mu + m_{FD}(x,t)\right)\psi, \quad (S8)$$

In this sense, the Floquet-Dirac mass term $m_{FD}(x,t)$ plays the two important roles of both the dynamical gauge field and Dirac mass term to realize the driven anomaly inflow on a domain wall. As compared with (S1), we start with a prototypical massive Floquet-Dirac Hamiltonian

$$H_{FD}(x,t) = -i\hbar\, \partial_x \sigma_1 + m(x,t)\sigma_3, \quad (S9)$$

with the time-periodic kink-loaded Floquet-Dirac mass term $m(x,t) = m \tanh x \cos\left(\frac{2\pi t}{T}\right)$. We calculated the Cauchy problem in Floquet-Dirac equation in terms of the Floquet periodicity (T), as shown in Fig. S3. For our calculations, we set the relevant parameters $\hbar = 1, v_F$ (or $c$) $= 1, m = 2$. The pi-mode domain wall in Fig. S3b, accompanying with the static domain wall in Fig. S3e, in the continuum limit, confirms the chief achievement based on the driven SSH lattice setup in the main text.

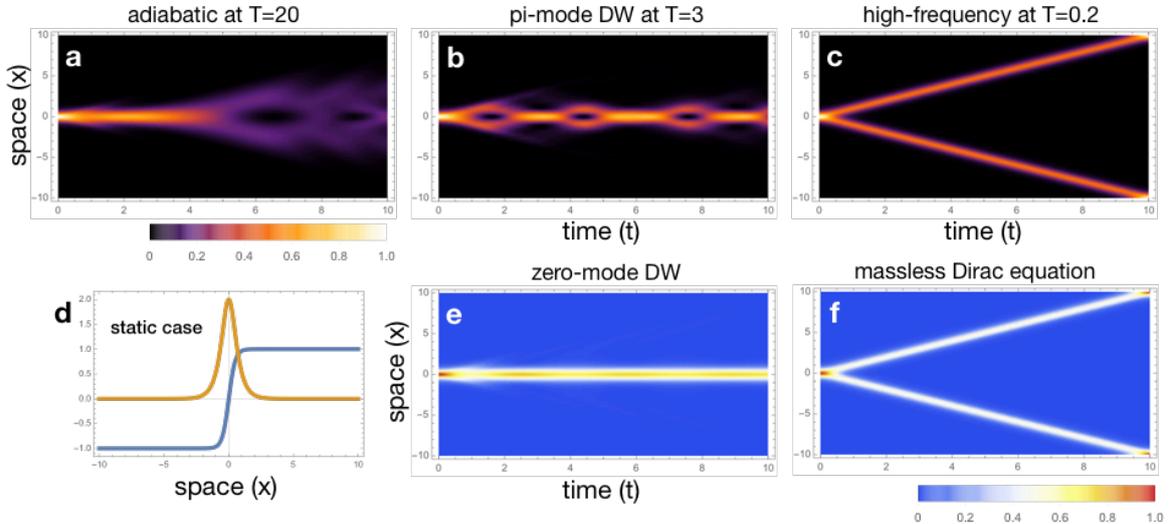

Fig. S3: (a-c) The Cauchy problem of Floquet-Dirac equation with time-periodic domain wall (DW) like mass term. (a) In adiabatic regime T=20; (b) In intermediate driven frequency regime T=3; (c) In high-frequency regime T=0.2. Only for properly-resonant driving frequencies, pi-mode DW then emerges. (d) As a calibration, we also solved the Cauchy problem with static kink-loaded mass term. (e) There exists the counterpart zero-mode DW as compared with the driven one. (f) The electron wavefunction propagates for massless Dirac equation, in comparison with the high-frequency case in (c).

## 4. Zero-mode anomaly on the domain wall (1+1D chiral anomaly)

To see the zero-mode anomaly problem intuitively, we can take the simple approach from the literature of Nielsen and Ninomiya 1983 [9], and the recent restatement in the excellent lecture of David Tong [3, 13]. Here, we will rather give a more concrete mathematical derivation from quantum field theory both in the IR and UV energy scales to directly reveal the connection between the Atiyah-Singer index theorem with the IR anomaly and the connection between the anomaly inflow mechanism with the UV anomaly. The 1+1 D chiral anomaly can be understood in terms of the Dirac sea (i.e., the vacuum, or ground state). We start from the Dirac equation

$$(i\gamma_\mu \partial^\mu + e\gamma_\mu A^\mu) = 0, \tag{S10}$$

Furthermore, we choose the 2-dimensional Dirac matrices in the following way

$$\gamma_0 = \sigma_2, \gamma_1 = i\sigma_1, \gamma_5 = \gamma_0\gamma_1 = \sigma_3. \tag{S11}$$

Then the Dirac equation is explicitly rewritten by

$$\left(i\frac{\partial}{\partial t} + \sigma_3\left(\frac{\partial}{\partial x} - eA_1\right)\right)\psi = 0, \tag{S12}$$

where $\sigma_3 \eta_\pm = \pm \eta_\pm$ corresponds to the left-handed (L, "+") and right-handed (R, "-") fermions. We study the above QED model with massless Dirac fermions with x-space compactified on a circle $S_1$ of length L. Then both the gauge fields and fermions are defined on a cylinder of space and time ($S_1 \times \mathbb{R}$) and obey the periodic (for bosons) and antiperiodic (for fermions) boundary conditions

$$A^\mu\left(t, x = -\frac{L}{2}\right) = A^\mu\left(t, x = \frac{L}{2}\right),$$
$$\psi\left(t, x = -\frac{L}{2}\right) = -\psi\left(t, x = \frac{L}{2}\right). \tag{S13}$$

Notice that we choose a gauge where $A_1$ is independent of x and $A_0 = 0$ (the Coulomb potential) can be neglected. Indeed, we treat $A_1(t)$ as an external electric field, which will be switched on adiabatically. Due to the periodicity in Eq. (S13), the reciprocal space of the gauge potential $A_1$ is a circle of length $2\pi/L$, in which the arbitrary values $A_1$ and $A_1 + 2\pi/L$ are gauge equivalent. According to periodic boundaries, the fermion wavefunction can be expanded into

$$\psi(t,x) = \frac{1}{\sqrt{L}} \sum_k u(k) e^{-iE_k t} \exp\left(i\frac{2\pi}{L}\left(k + \frac{1}{2}\right)x\right), \tag{S14}$$

where $u(k)$ is the Bloch component and the integer $k = 0, \pm 1, \pm 2 \ldots$ Substituting the trial function (S14) in the Dirac equation (S12), we obtain the following energy spectrum for the L- and R- fermions

$$E_k^{(\pm)} = \pm\left[\frac{2\pi}{L}\left(k + \frac{1}{2}\right) + A_1\right]. \tag{S15}$$

So, the energy spectrum is comfortably discrete because of the periodicity, and it depends linearly on the gauge potential $A_1$ along the x-direction, as shown in Fig. S4.

Now we turn from the single-particle excitation picture. At $A_1 = 0$ we have a ground state – the Dirac sea – all the fermion eigenstates with negative energies ($E_k^{(\pm)} < 0$) are infinitely filled (see the blue circles in Fig. S4a). However, if we increase $A_1$ from 0 to $2\pi/L$ we create a L-particle and a R-antiparticle (i.e., hole). Thus, we obtain the charges

$$\Delta Q = \Delta \int dx\, j_0(t,x) = \Delta Q_L + \Delta Q_R = 1 - 1 = 0,$$
$$\Delta Q^5 = \Delta \int dx\, j_0^5(t,x) = \Delta Q_L - \Delta Q_R = 1 + 1 = 2. \tag{S16}$$

This change compared with the increase of gauge potential $\Delta A_1 = 2\pi/L$ and per time unit ($\Delta t$) gives

$$\frac{\Delta Q^5}{\Delta t} = \frac{L}{\pi} \frac{\Delta A_1}{\Delta t}.$$

Considering the infinite local change in the equation

$$\frac{\partial}{\partial t} \int_0^L dx\, j_0^5(t,x) = \frac{1}{\pi} \frac{\partial}{\partial t} \int_0^L dx\, A_1(t).$$

Finally, we arrive at the 1+1 D chiral anomaly

$$\partial_0 j_0^5 = \frac{1}{\pi} \partial_0 A_1 \xrightarrow{yields} \partial^\mu j_\mu^5 = \frac{1}{\pi} \epsilon^{\mu\nu} \partial_\mu A_\nu. \tag{S17}$$

where the axial current in QFT description is defined by $j_\mu^5 = \bar\psi \gamma_\mu \gamma_5 \psi$.

Alternatively, we can study the behavior at some UV cutoff, which is directly related to the perturbative approach in QFT. We have to face the emergence of infinity when dealing with the Dirac sea because the total energy and total charge of the vacuum are ill-defined and diverge. In the procedure of QFT [2, 8], the point-splitting method can regularize the currents and meanwhile preserves gauge invariance, leading to the vector and axial currents

$$\begin{aligned}
j_\mu^{reg}(t,x) &= \lim_{\varepsilon \to 0} \bar\psi(t, x+\varepsilon) \gamma_\mu \psi(t,x) \exp\left(-i \int_x^{x+\varepsilon} dx\, A_1\right), \\
j_\mu^{5reg}(t,x) &= \lim_{\varepsilon \to 0} \bar\psi(t, x+\varepsilon) \gamma_\mu \gamma_5 \psi(t,x) \exp\left(-i \int_x^{x+\varepsilon} dx\, A_1\right).
\end{aligned} \tag{S18}$$

Correspondingly, the regularized charges are defined by

$$\begin{aligned}
Q &= \int dx\, j_0^{reg}(t,x) = Q_L^{reg} + Q_R^{reg}, \\
Q^5 &= \int dx\, j_0^{5reg}(t,x) = Q_L^{reg} + Q_R^{reg}.
\end{aligned} \tag{S19}$$

in which the axial charge for L- and R-particle is obtained explicitly from Eq.(S14),

$$Q_{L,R}^{\text{reg}} = \lim_{\varepsilon \to 0} \int dx \, \bar{\psi}_{L,R}(t, x+\varepsilon) \gamma_0 \psi_{L,R}(t, x) \exp\left(-i \int_x^{x+\varepsilon} dx A_1\right)$$
$$= \lim_{\varepsilon \to 0} \int dx \, \psi_{L,R}^\dagger(t, x+\varepsilon) \psi_{L,R}(t, x) \exp(-i\varepsilon A_1) \quad (S20)$$
$$= \sum_{k \in \text{filled}} \exp\left(-i\varepsilon \left[\frac{2\pi}{L}\left(k_{L,R} + \frac{1}{2}\right) + A_1\right]\right)$$

with the momentum indices $k_L = -1, -2, \ldots; k_R = 0, 1, 2, \ldots$ If we first take the limit $\varepsilon \to 0$, we return to the unregularized ill-defined currents, e.g.,

$$Q^5 = Q_L - Q_R = \sum_{k_L \in \text{filled}} 1 - \sum_{k_R \in \text{filled}} 1 = \infty - \infty, \quad (S21)$$

corresponding to the infinity difference in the Dirac sea, which connects the index of the Dirac operator from the Atiyah-Singer index theorem. To avoid the divergence, therefore, we fix the regulator as a finite quantity and finish the summation in (S20), we obtain

$$Q_L^{\text{reg}} = -\frac{L}{2\pi} \frac{1}{i\varepsilon} + \frac{L}{2\pi} A_1,$$
$$Q_R^{\text{reg}} = +\frac{L}{2\pi} \frac{1}{i\varepsilon} - \frac{L}{2\pi} A_1. \quad (S22)$$

Now, we obtain the regularized currents

$$Q = Q_L^{\text{reg}} + Q_R^{\text{reg}} = 0,$$
$$Q^5 = Q_L^{\text{reg}} - Q_R^{\text{reg}} = -\frac{L}{\pi} \frac{1}{i\varepsilon} + \frac{L}{\pi} A_1. \quad (S23)$$

We address two aspects here. First, the vector current is conserved so that the gauge invariance ($\psi(t, x) \to e^{i\alpha(x)} \psi(t, x)$) is preserved at the quantum level. Second, the axial charge is interesting that contains divergent constant ($1/\varepsilon$), which can be subtracted to define the regularized current, because of the independence of the external gauge potential $A_1$. As we increase $A_1$ from 0 to $2\pi/L$. As a result, the physical axial charge changes by

$$\Delta Q^5 = \frac{L}{\pi}\left(\frac{2\pi}{L}\right) = 2, \quad (S24)$$

providing the same chiral anomaly result as presented in (S16, S17). Now, we find that at the quantum level, the axial current is not conserved due to the anomalous current contribution, and therefore, resulting in violating the chiral gauge transformation ($\psi(t,x) \to e^{i\alpha(x)\gamma_5}\psi(t,x)$).

In the above derivation, we provide two different perspectives to understand the emergence of the chiral anomaly from the infinite Dirac sea: the anomaly arises either as an IR phenomenon in the infra-red energy scale (S16)) or as a UV-phenomena in the ultra-violet energy scale (S23). This matching between IR and UV behaviors in mind also offers us an intuitive understanding of the Callan-Harvey anomaly inflow mechanism that cancels the zero-mode anomaly on the domain wall (IR-anomaly) by the topological bulk's boundary term contribution (UV-anomaly). For instance, consider that there is a chiral zero mode ($E = p_y$) traveling along with the interface on the 2+1D domain wall, we can expect that in the presence of an external electromagnetic field $A_\mu$, the chiral fermion has a gauge anomaly

$$\partial^\mu j_\mu^{(L)} = \frac{1}{2\pi}\epsilon^{\mu\nu}\partial_\mu A_\nu \to \dot{Q}^{(L)} = \frac{L}{2\pi}E_y, \tag{S25}$$

where the electric component is given by $E_y = -\frac{\partial}{\partial t}A_1$. The ½-factor is contributing to the absence of the right-handed zero modes on the wall. The gauge anomaly can be rephrased as an additional contribution to the action under gauge transformation $A_\mu \to A_\mu + \partial_\mu \alpha(x)$. To be precise

$$\Delta S_\alpha = \frac{1}{2\pi}\int d^2x \alpha(x)\epsilon^{\mu\nu}\partial_\mu A_\nu. \tag{S26}$$

Due to the missing contribution of the right-handed fermion, the gauge charge of the rest left-handed fermion is not conserved anymore. There must be something in the massive degree of freedom in the bulk, which exactly cancels this contribution and restores gauge invariance. This motivation is associated with an inflow of charge from the extra dimension (i.e., the bulk).

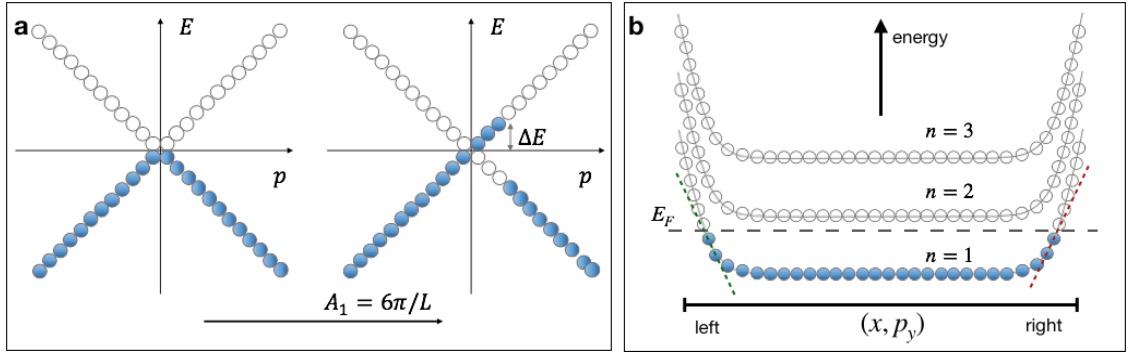

Fig. S4: (a)The energy spectrum of the chiral fermions. The left case at $A_1 = 0$ presents the Dirac sea at equilibrium, while the right case by adiabatically increasing the gauge field $A_1 \neq 0$ presents the shifted energy spectrum. (b) The schematic chiral edge-state picture of the 2D quantum Hall effect with only the first Landau level filled (n=1). The Landau levels are lifted up at edges due to the potential confinement.

## 5. Anomaly inflow mechanism in QHE heterostructure (from the bulk)

To reveal the anomaly inflow in quantum Hall effect, we consider a Chern-Simons form in 2-dimensional QHE with a heterostructure configured the external magnetic field (B), as shown in Fig. 3a of the main text. The known Chern-Simons action is then given by

$$S_{CS}[A] = \frac{k}{4\pi} \int d^3 x \epsilon^{\mu\nu\lambda} A_\mu \partial_\nu A_\lambda. \qquad (S27)$$

The coefficient k is called the level of the Chern-Simons term. Under a gauge transformation $A_\mu \to A_\mu + \partial_\mu \alpha$,

$$S_{CS}[A] \to S_{CS}[A] + \frac{k}{4\pi} \int d^3 x \partial_\mu \left( \alpha \epsilon^{\mu\nu\lambda} A_\mu \partial_\nu A_\lambda \right). \qquad (S28)$$

The additional term is a total derivative that contributes to a non-trivial boundary term. First, we can quickly compute the current that arises from the Chern-Simons action, which is given by

$$J_i = \frac{\delta S_{CS}[A]}{\delta A_i} = -\frac{k}{2\pi} \varepsilon_{ij} E_j \to J_x = \sigma_{xy} E_y, \qquad (S29)$$

with a Hall conductivity

$$\sigma_{xy} = \frac{k}{2\pi} = \frac{ne^2}{h}. \tag{S30}$$

This matches the Hall current of $n$ filled Landau levels if we identify the Chern-Simons level with $k = \frac{ne^2}{\hbar}$. Consider an open boundary condition in the x-direction. The confining potential pushes up Landau levels at the edges of the sample, see the edge state picture of QHE in Fig. S4b. The Fermi energy ($E_F$) will cross Landau levels at the edges and lead to the metallic edge state, but the bulk state is still insulating. Interestingly, the one-dimensional edge states along the y-direction are chiral in terms of the left- and right-handed boundaries. Now we have known that on edge there is a gauge anomaly (S17, S25)

$$\dot{\rho}^{(L)} = \frac{k}{2\pi} E_y, \dot{\rho}^{(R)} = -\frac{k}{2\pi} E_y, \tag{S31}$$

where the charge density $\rho^{(L,R)} = Q^{(L,R)}/L$ with the length L in the y-direction. It is straightforward to observe that a uniform electric field ($E_y$) induces a charge current along the wall, which is exactly canceled by the current form Chern-Simons Hall current ($J_x$) that flows towards the edge. More precisely, for the QHE heterostructure at Landau level $n = 1$, the zero-mode anomaly on the domain wall is

$$\dot{\rho} = \frac{2e^2}{h} E_y, \tag{S32}$$

which matches the summation of Hall currents from both sides $J_x + J_x = 2\frac{e^2}{h} E_y$, as shown in Fig. 3a of the main text. This is known as the Callan-Harvey anomaly inflow mechanism.

## 6. Numerical simulations of the time-periodic SSH model

Under the open boundary condition, the Hamiltonian of the domain walls in the static/driven SSH model can be easy expressed in the matrix form of

$$H = \begin{pmatrix} 0 & \kappa_0 - \delta\kappa_1 \cos(\omega t + \theta^{(L)}) & 0 & 0 & 0 & 0 & 0 & \cdots \\ \kappa_0 - \delta\kappa_1 \cos(\omega t + \theta^{(L)}) & 0 & \kappa_0 + \delta\kappa_1 \cos(\omega t + \theta^{(L)}) & 0 & 0 & 0 & 0 & \cdots \\ 0 & \kappa_0 + \delta\kappa_1 \cos(\omega t + \theta^{(L)}) & 0 & \ddots & \ddots & \ddots & \ddots & \cdots \\ 0 & 0 & \ddots & 0 & \kappa_0 - \delta\kappa_1 \cos(\omega t + \theta^{(L)}) & 0 & 0 & \cdots \\ 0 & 0 & 0 & \kappa_0 - \delta\kappa_1 \cos(\omega t + \theta^{(L)}) & 0 & \kappa_0 - \delta\kappa_1 \cos(\omega t + \theta^{(R)}) & 0 & \cdots \\ 0 & 0 & 0 & 0 & \kappa_0 - \delta\kappa_1 \cos(\omega t + \theta^{(R)}) & 0 & \kappa_0 + \delta\kappa_1 \cos(\omega t + \theta^{(R)}) & \cdots \\ 0 & 0 & 0 & 0 & 0 & \kappa_0 + \delta\kappa_1 \cos(\omega t + \theta^{(R)}) & 0 & \ddots \\ \vdots & \vdots & \vdots & \vdots & \vdots & \vdots & \ddots & \end{pmatrix}_{(2N+1)\times(2N+1)}$$

(S33)

In the adiabatic limit, the instantaneous energy spectrum can be calculated directly by diagonalization, see Fig. 2a, Fig. S5, S6. On the other hand, to get the quasi-energy spectrum in a full driven frequency range and the time-dependent evolution of the system, we consider the evolution operator

$$|\psi(t)\rangle = U(t, t_0)|\psi(t_0)\rangle, \tag{S34}$$

where $|\psi(t)\rangle$ denotes the state of the system at time $t$ and $|\psi(t_0)\rangle$ denotes the initial state at time $t_0$. The time evolution operator $U(t, t_0)$ is determined by the differential equation

$$i\partial_t U(t, t_0) = H(t)U(t, t_0), \tag{S35}$$

which has the solution

$$U(t, t_0) = \hat{T} e^{-i\int_{t_0}^{t} H(t')dt'}, \tag{S36}$$

with an initial value $U(t, t_0) = Id$ and $\hat{T}$ being the time-ordering operator. If the system has the discrete time-translation symmetry (with periodicity $T$), then the evolution operator satisfies $U(t_2, t_1) = U(t_2 + T, t_1 + T)$ and $U(t_2, t_1) = U(t_2, t')U(t', t_1)$ for arbitrary time $t'$ in between the interval $(t_1, t_2)$. The dynamic evolution operator from time $t_0$ to $t$, can be rewritten as

$$U(t, t_0) = U(t, t_0 + nT)U(t_0 + nT, t_0) = U(t, t_0 + nT)[U(t_0 + T, t_0)]^n, \tag{S37}$$

where n is an integer and the time interval within one cycle $\delta t = (t - t_0 + nT) \in [0, T]$. From (S37), we can see that the stroboscopic evolution observed at each period can be fully described by $U(t_0 + T, t_0)$, called as the Floquet operator, which satisfies

$$U(t_0 + T, t_0)|\psi(t_0)\rangle = e^{-i\varepsilon T}|\psi(t_0)\rangle, \tag{S38}$$

where $|\psi(t_0)\rangle$ is the Floquet state and $\varepsilon$ is the quasienergy. It is convenient to define the evolution within one period as an evolution with the time-independent effective Hamiltonian

$$U(t_0 + T, t_0) = e^{-iH_F[t_0]T}, \tag{S39}$$

where the choice of $t_0$ is arbitrary, which is called the Floquet gauge. In this case, the effective Hamiltonian is defined by

$$H_F[t_0] = \frac{i}{T}\log_{-\eta} U(t_0 + T, t_0), \tag{S40}$$

where the function $log_\eta$ is the complex logarithm with branch cut along an axis with angle η [34], defined as

$$log_{-\eta} e^{i\varphi} = i\varphi, \quad \text{for} \quad -\eta - 2\pi < \varphi < -\eta. \tag{S41}$$

Then we can obtain the quasienergy spectrum of the effective Hamiltonian as exhibited in Fig. 2b of the main text. The dynamic evolution of the system can be calculated numerically by the discretized evolution operator

$$U(t,0) = \lim_{\Delta t \to 0} e^{-iH(t-\Delta t)\Delta t} e^{-iH(t-2\Delta t)\Delta t} \ldots e^{-iH(\Delta t)\Delta t} e^{-iH(0)\Delta t}, \tag{S42}$$

The results are presented in Fig. 3b, Fig. 4d-4f, with an input wavefunction from the middle site of the waveguide array to excite the domain wall localized states.

Under the periodic boundary condition, the Hamiltonian of the SSH model can be rewritten in the momentum space as

$$H(k,t) = \big((\kappa_0 - \delta\kappa_1 \cos(\omega t)) + (\kappa_0 + \delta\kappa_1 \cos(\omega t))\cos(k)\big)\sigma_x + (\kappa_0 + \delta\kappa_1 \cos \omega t)\sin(k)\,\sigma_y \tag{S43}$$

where $\sigma_x$, $\sigma_y$ are the Pauli matrices in the basis of sublattices A and B. We notice that this Hamiltonian has chiral symmetry which is defined as the unitary chiral operator $\Gamma = \sigma_z$

$$\Gamma H(t,k)\Gamma^{-1} = -H(-t,k) \tag{S44}$$

and also, the discrete time-translation symmetry

$$H(k,t) = H(k, t+T). \tag{S45}$$

A $\mathbb{Z}_2$-valued bulk gap invariant $\nu_\epsilon[U] \in \mathbb{Z}_2$ defined for the chiral gaps $\pi$ can work in the driven system with the novel chiral symmetry. Note that the chiral symmetry has the constraint on the periodized evolution operator (micromotion dynamics)

$$\Gamma V_\varepsilon(t,k)\Gamma^{-1} = -V_{-\varepsilon}(-t,k)e^{2\pi i t/T} \tag{S46}$$

For $\epsilon = \pi$ and $t = T/2$, it exactly turns out [31,34]

$$\Gamma V_\pi(T/2,k)\Gamma^{-1} = V_\pi(T/2,k) \tag{S48}$$

which is diagonal in the chiral basis.

$$V_\pi(T/2,k) = \begin{pmatrix} V_\pi^+ & 0 \\ 0 & V_\pi^- \end{pmatrix}. \tag{S49}$$

The chiral invariant is defined by

$$\nu_\pi = \frac{i}{2\pi}\int_{-\pi}^{\pi} tr((V_\pi^+)^{-1}\partial k\, V_\pi^+)dk. \tag{S50}$$

The $\pi$ mode emerges in the frequency region $1/3 < \omega/\Delta < 1$ with the nontrivial $\pi$-gap invariant $\nu_\pi = 1$, as shown in Fig. 2b of the main text.

# 7. Additional experimental data of pi modes and pi domain walls in the periodic-driven TI/NI heterostructures

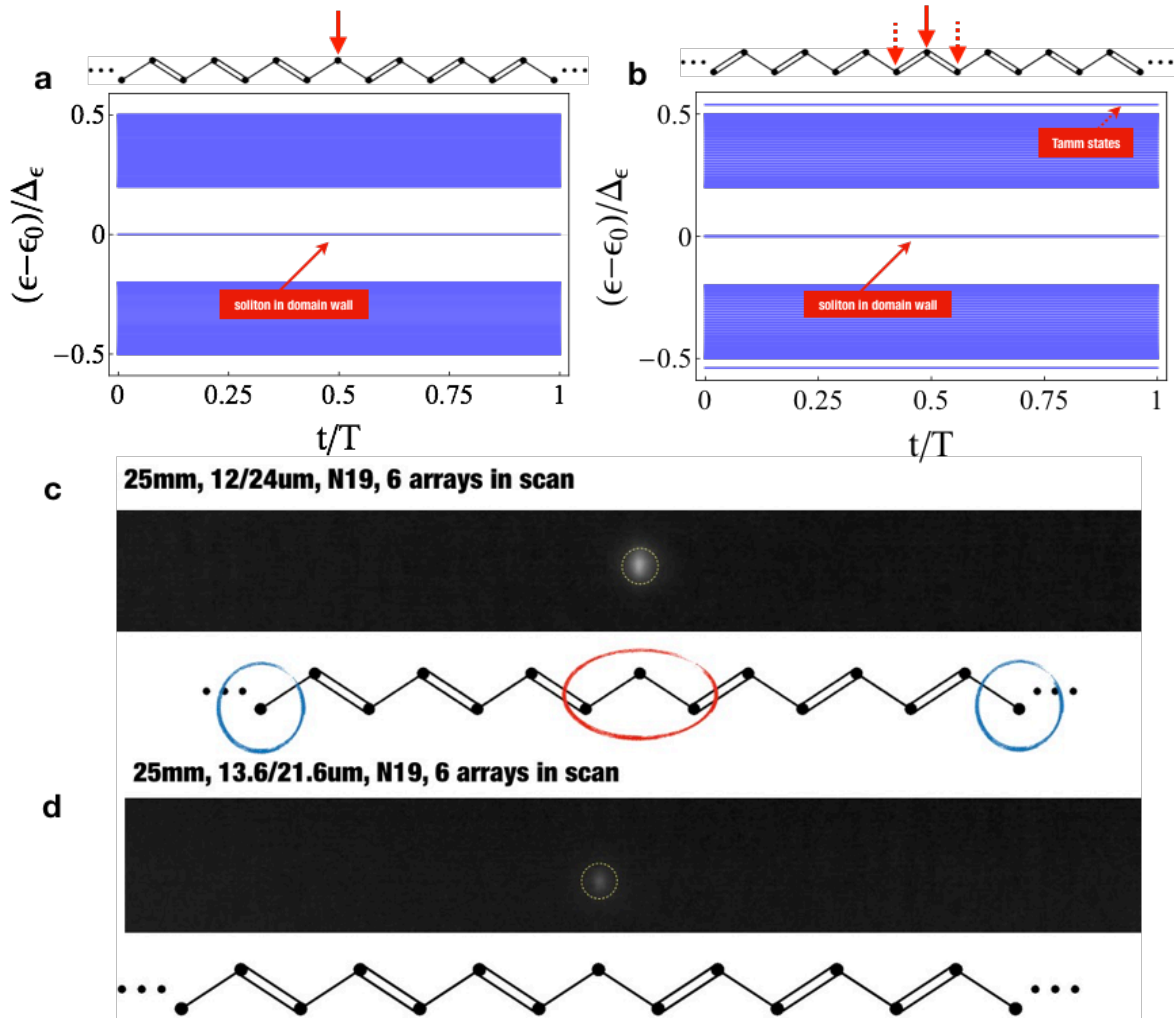

Figure S5: (a) The soliton excitation in domain wall with kink structure (two-weak-bond coupling profile). (b)The soliton excitation in domain wall with anti-kink structure (two-strong-bond coupling profile). (c-d) We made a video to present the scanning process of the input, see the videos in the attached file.

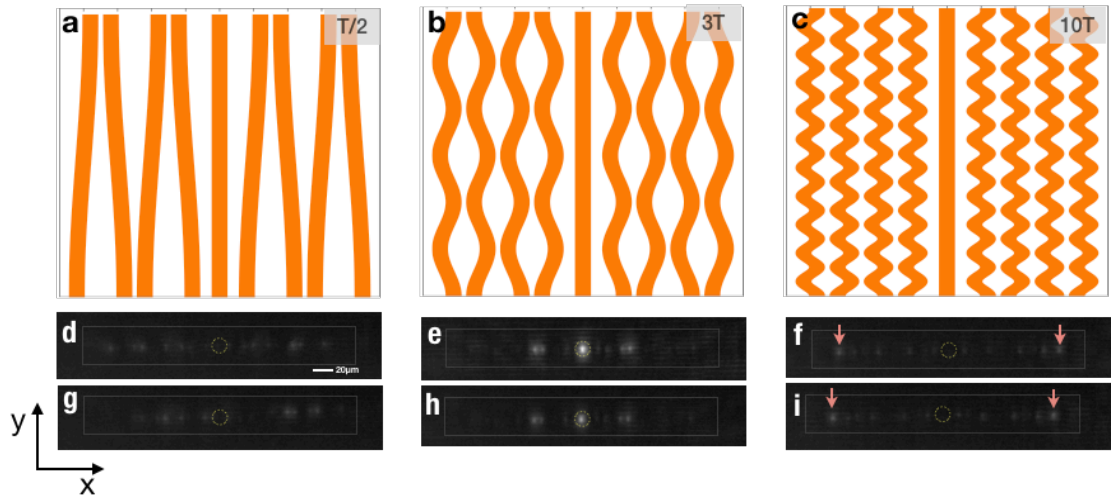

Fig. S6: Additional experimental observations of driven domain wall dynamics ranging from adiabatic limit to high-frequency limit, as a supplementary figure for Fig. 4 in the main text. The parameters are same as in the main text: 25mm/3T, 10/20um, N19.

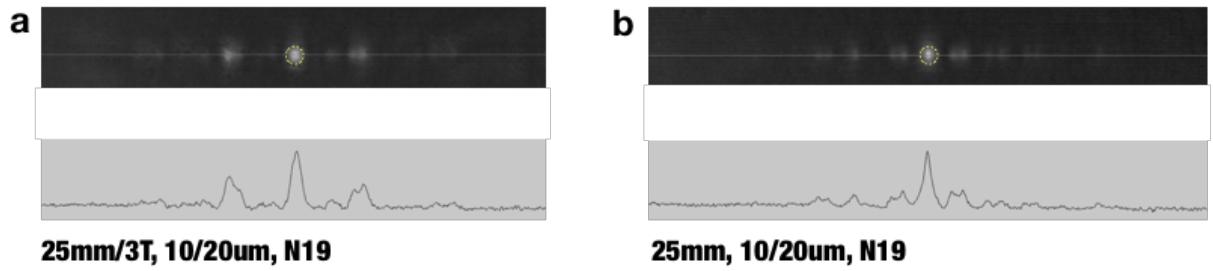

Fig. S7: Comparison between Floquet domain walls and static domain wall in the SSH array setup. Both two output intensities show the localization properties in the domain wall configurations of the photonic arrays.

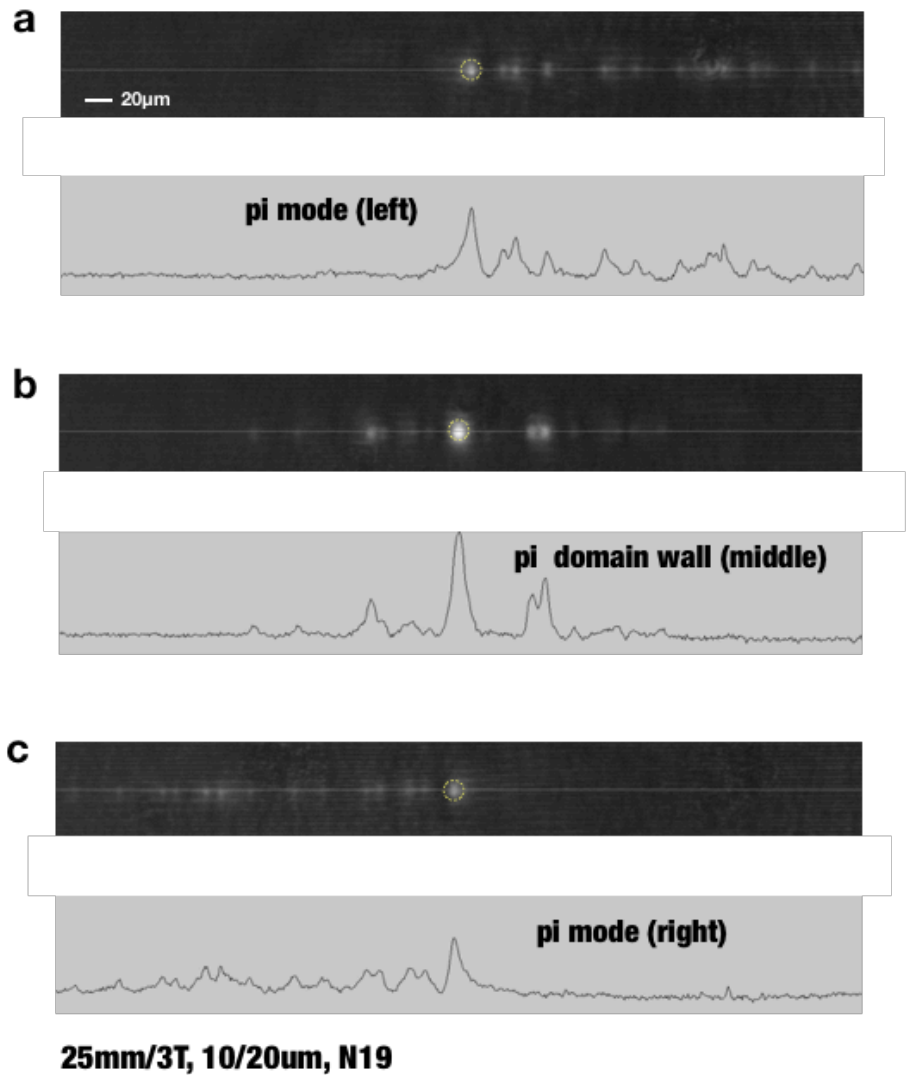

Fig.S8: The output distribution of pi modes at the two ends of the driven SSH chain and pi domain wall in the middle. The circles mark the input waveguides, that are, the left end waveguide (the 1st waveguide), the middle waveguide (the 10th waveguide) and the right end waveguide (the 19th waveguide)